\documentclass[a4paper,UKenglish,cleveref, autoref, thm-restate]{lipics-v2021}
\pdfoutput=1 %uncomment to ensure pdflatex processing (mandatatory e.g. to submit to arXiv)
\hideLIPIcs  %uncomment to remove references to LIPIcs series (logo, DOI, ...), e.g. when preparing a pre-final version to be uploaded to arXiv or another public repository

\title{Online Bin Packing with Item Size Estimates} %TODO Please add

\author{Matthias {Gehnen}}{RWTH Aachen University, Germany  \and \url{https://tcs.rwth-aachen.de/users/gehnen/} }{gehnen@cs.rwth-aachen.de}{https://orcid.org/0000-0001-9595-2992}{Supported by a Research ENHANCE.R grant by RWTH Aachen}%TODO mandatory, please use full name; only 1 author per \author macro; first two parameters are mandatory, other parameters can be empty. Please provide at least the name of the affiliation and the country. The full address is optional. Use additional curly braces to indicate the correct name splitting when the last name consists of multiple name parts.

\author{Andreas Usdenski}{Paderborn University, Germany}{andreas.usdenski@uni-paderborn.de}{[orcid]}{}

\authorrunning{M. Gehnen and A. Usdenski} %TODO mandatory. First: Use abbreviated first/middle names. Second (only in severe cases): Use first author plus 'et al.'

\Copyright{Matthias Gehnen and Andreas Usdenski} %TODO mandatory, please use full first names. LIPIcs license is "CC-BY";  http://creativecommons.org/licenses/by/3.0/

 %TODO mandatory: Please choose ACM 2012 classifications from https://dl.acm.org/ccs/ccs_flat.cfm 

 %TODO mandatory; please add comma-separated list of keywords

\category{} %optional, e.g. invited paper

\relatedversion{} %optional, e.g. full version hosted on arXiv, HAL, or other respository/website
\acknowledgements{}%optional

\nolinenumbers %uncomment to disable line numbering

\EventEditors{}
\EventNoEds{2}
\EventLongTitle{}
\EventShortTitle{}
\EventAcronym{}
\EventYear{}
\EventDate{}
\EventLocation{}
\EventLogo{}
\SeriesVolume{}
\ArticleNo{}
\usepackage{pgfplots}
\usepgfplotslibrary{fillbetween}
\pgfsetlayers{background,edgelayer,nodelayer,main}
\usepackage{algorithm}
\usepackage{algorithmicx}
\usepackage[noend]{algpseudocode}
\usepackage{rotating}

\begin{document}

\maketitle

\begin{abstract}
Imagine yourself moving to another place, and therefore, you need to pack all of your belongings into moving boxes with some capacity. In the classical bin packing model, you would try to minimize the number of boxes, knowing the exact size of each item you want to pack. In the online bin packing problem, you need to start packing the first item into a box, without knowing what other stuff is upcoming. 

Both settings are somewhat unrealistic, as you are likely not willing to measure the exact size of all your belongings before packing the first item, but you are not completely clueless about what other stuff you have when you start packing. In this article, we introduce the \emph{online bin packing with estimates} model, where you start packing with a rough idea about the upcoming item sizes in mind.  

In this model, an algorithm receives a size estimate for every item in the input list together with an accuracy factor $\delta$ in advance. Just as for regular online bin packing the items are then presented iteratively. The actual sizes of the items are allowed to deviate from the size estimate by a factor of $\delta$. Once the actual size of an item is revealed the algorithm has to make an irrevocable decision on the question where to place it. This is the first time online bin packing is studied under this model.

This article has three main results: First, no algorithm can achieve a competitive ratio of less than $\frac{4}{3}$, even for an arbitrary small factor $\delta>0$. Second, we present an algorithm that is $1.5$-competitive for all $\delta \leq \frac{1}{35}$. Finally, we design a strategy that yields a competitive ratio of $\frac{4}{3}$ under the assumption that not more than two items can be placed in the same bin, which is best possible in this setting.
\end{abstract}

\newpage
\section{Introduction}
Many optimization problems in theoretical computer science can be studied in a version where the entire input instance is not shown in advance but instead revealed piece-by-piece and an algorithm has to make decisions based on that limited information. Such problems are called \textit{online problems} and the corresponding algorithms \textit{online algorithms}. Of course, it might not always be possible to achieve the same cost or gain in this setting compared to a strategy that knows the entire input instance in advance. For this reason, the main goal when designing online algorithms is to get as close to the optimal solution in terms of cost or gain as possible.

One problem that has been extensively studied in its online version is \textit{Bin Packing}. In this problem, a set of items has to be packed in as few equal-sized bins as possible. It is well known that the offline version of bin packing is computationally difficult to solve exactly, however, a asymptotic polynomial-time approximation scheme exists \cite{fernandez1981bin,hoberg2017logarithmic}. Also, the online variant performs quite well: Even the simplest conceivable strategy achieves a packing that uses at most twice as many bins as the optimal solution for any input instance. At the same time, it is also known that there exists no algorithm that can pack every possible set of items in a way such that the ratio between the number of bins it uses and the optimal solution (also known as the competitive ratio of the algorithm) is lower than $1.54$ \cite{balogh2021new}.

Online bin packing algorithms have no information on the future and always have to factor in the possibility that the input instance might end at any moment. In practice, however, it is often reasonable to assume that one knows how many items have to be packed in advance and even how large these items roughly are. The estimates might not be perfect if, for example, the sizes of the items are simply measured "by eye" or the input data is affected by numerical errors. Intuitively, this additional information should help to pack the items more effectively, assuming that the predictions are not too far off. But how much can this information improve the results compared to the standard online model, and how dependent are they on the quality of the estimates?

In this article, we introduce a new formal model for online bin packing with estimated item sizes. In this model, an algorithm is given additional information on the input instance before any items are shown. This information consists of an estimate of the size of every item in the list and an accuracy factor $\delta$. The actual sizes of the items are contained in an interval determined by the estimate together with the accuracy factor. To the best of our knowledge, this is the first time online bin packing has been studied under this model.

We analyze upper and lower bounds on the asymptotic competitive ratio of the problem depending on the prediction error $\delta$. The most notable findings are: First, no algorithm can achieve a competitive ratio of less than $\frac{4}{3}$ for any accuracy $\delta$. We also show that for $\delta > \frac{41}{43}$ the competitive ratio is bounded by $1.5$ from below. Second, we design an algorithm that is $1.5$-competitive for every $\delta \leq \frac{1}{35}$. Finally, we present a strategy that yields a competitive ratio of $\frac{4}{3}$, if not more than items fit into one bin for all $\delta > 0$, and might also be applied in related variants of online bin packing.

This article is structured as follows: First, we review the most important results for regular online bin packing as well as other important work related to our variant. We then start by formally defining bin packing with estimated item sizes.

The lower bounds of this article can be found in Section 2. We first establish a lower bound for precise estimates with an arbitrary $\delta > 0$ and then move on to analyzing the case of more imprecise estimates. Here we first show a lower bound for a medium-sized allowed deviation, before closing the section by showing that there cannot be a $1.5$-competitive algorithm if $\delta$ is sufficiently large.

In Section 3, we present the \textsc{Planned-Harmonic} algorithm and show that it is $1.5$-competitive for every $\delta \leq \frac{1}{35}$. In Section 4 we first discuss some of the previous attempts to deal with instances where no three items can be packed together, before introducing our approach called \textsc{Delayed-Best-Fit}. Finally, we raise some open questions and discuss the insights of this article.
\subsection{Related Work}

Formally, the setting of online algorithms and the competitive analysis, as we are performing in this article, was introduced in 1985 by Sleator and Tarjan\cite{10.1145/2786.2793}. Note that however online problems have been studied before, for example, the here considered online bin packing problem. We refer to the books by Borodin and El-Yaniv \cite{borodin2005online} and Komm \cite{komm2016introduction} for a general overview of online algorithms and competitive analysis.

For the online bin packing problem, several strategies are studied about how to deal with upcoming items. It is well known that already very simple strategies can provide good competitive ratios, that might already be sufficient in some applications: The strategy \textsc{Next Fit}, where each item is just placed in the last used bin (if it fits there) or otherwise placed in a new bin, is already known to be $2$-competitive \cite{johnson1973near}. If each new item is placed in the bin which is fullest among those in which the new item fits (and if it fits in none, it will be placed in a new bin), it is known as \textsc{Best Fit} and improves the competitive ratio to $1.7$ \cite{johnson1974worst}. The same ratio can also be achieved, when each new item is placed in the first bin it fits into, following a strategy called \textsc{First Fit}\cite{johnson1974worst}.

This upper bound was beaten by Yao \cite{yao1980new}, who introduced an algorithm called \textsc{Refined-First-Fit}. This strategy attempts to pack items with a size larger than $\frac{1}{3}$ more efficiently by dividing them into subclasses based on their size and preferring to pack items of certain subclasses together. It yields a competitive ratio of $\frac{5}{3}$ and thus performs significantly better than the simple strategies discussed above.

Further improvements of the competitive ratio usually rely on the concept of \textsc{Harmonic} Algorithms. Here, each item will be classified by its size; and, depending on its class, treated differently \cite{lee1985simple}. 
In its simple version, first, a constant $M \geq 3$ is fixed. Then, an item is classified to be an $I_k$ item, if its size is in in the interval $(\frac{1}{k + 1}, \frac{1}{k}]$ for $k \leq M-1$. Items, which are smaller than $\frac{1}{M}$ are classified to be an $I_M$ item.
Items of each class are packed together greedily, meaning each bin only contains items from the same class.
The competitive ratio of an algorithm $\textsc{Harmonic}_M$ decreases with increasing $M$, and converges towards $1.691$ as $M$ goes to infinity. The ratio of \textsc{Best-Fit} however ($1.7$) is already beaten for $M =7$ \cite{lee1985simple}.

Lee and Lee \cite{lee1985simple} combined a similar idea from \textsc{Refined-First-Fit}  with $\textsc{Harmonic}_M$ to create an algorithm called \textsc{Refined-Harmonic} which achieves a ratio of $\frac{373}{228} \approx 1.6369$. 
Most approaches that improve these results use $\textsc{Harmonic}_M$ as their basis. The first improvements were presented by Ramanan et al. \cite{ramanan1989line} who introduced the algorithms \textsc{Modified-Harmonic} and \textsc{Modified-Harmonic-2} with competitive ratios of around $1.6156$ and $1.612$, respectively. Later, Seiden \cite{seiden2002online} managed to generalize multiple previously known algorithms to a strategy he called \textsc{Super-Harmonic} and thereby showed that all these algorithms can be analyzed similarly. He also introduced a new algorithm called \textsc{Harmonic++} based on \textsc{Super-Harmonic} and achieves a competitive ratio of $1.58889$. The currently best-known strategy for online bin packing was proposed in \cite{balogh2017new} and yields a ratio of approximately $1.57829$.

However, one cannot expect to decrease the competitive ratio much more: Yao proved that every algorithm for the online bin packing problem has to be at least $1.5$-competitive \cite{yao1980new}. Brown and Liang \cite{brown1979lower}\cite{liang1980lower} independently improved on this lower bound by adding more item sizes to the instance described above, obtaining a new lower bound of around $1.5363$. This value was improved to $1.54015$ by van Vliet \cite{van1992improved} and later to first $1.54037$ and then $1.54278$ by Balogh et al. \cite{balogh2012new,balogh2021new}. This is the currently best lower bound on the online bin packing problem.

An extensive survey on upper and lower bounds for online bin packing together with some of its variants can be found in \cite{coffman2013bin}.

\subsection{Relaxed online computation}
There has also been research on scenarios where the algorithm is given additional information about the input instance before any items are revealed. One of these settings is the advice model. Here, the algorithm has access to an additional advice tape which an oracle with perfect knowledge on the input instance can write arbitrary information on. The upper and lower bounds on the competitive ratio that can be determined in this setting are dependent on the number of bits the algorithm is allowed to read on the advice tape. An overview of the advice model can be found in Komm \cite{komm2016introduction} and in the survey by Boyar et al. \cite{boyar2017online}.

A simple heuristic that yields a competitive ratio of $\frac{3}{2}$ and requires $O(\log n)$ advice bits ($n$ being the number of items in the input list) is called \textsc{Reserve-Critical} and was introduced by Boyar et al. in \cite{boyar2016online}. Here, the advice is used to tell the algorithm the number of items with a size in the interval $(\frac{1}{2}, \frac{2}{3}]$. The algorithm then reserves a bin for each of these items and tries to pack them with smaller ones. A major improvement on this result was given by Angelopoulos et al. \cite{angelopoulos2018online} who showed that it is even possible to beat $\frac{3}{2}$ with a constant number of advice bits; their algorithm \textsc{DR+} yields a competitive ratio of around $1.47012$. In the same paper, they also observed that it only takes as little as $16$ bits for an algorithm to outperform every algorithm for classical online bin packing. Renault et al. showed that there exists an algorithm that can construct an optimal packing using only linear advice \cite{renault2015online}. Finally, it is known that any algorithm needs to be able to access an advice string with a length of at least $\Omega(n)$ to beat a competitive ratio of $4 - 2\sqrt{2} \approx 1.172$ \cite{mikkelsen2015randomization}.

Another approach is taken by Angelopoulos et al. \cite{angelopoulos2021online} which combines the standard online bin packing problem with predictions (also known as machine-learned advice). The paper assumes that the item sizes are drawn from a fixed set of integers, and the predictions are used to predict the frequency at which different sizes occur in the instance. In contrast to the advice model, the predictions are prone to some level of distortion quantified by a parameter that the algorithm does not know. The prediction model was introduced by Purohit et al. \cite{purohit08}, has become quite popular in recent years, and was also applied to several other problems. An extensive overview of the most important results for online bin packing under the advice and prediction models is given by Kamali\cite{kamali2020online}; for more on machine-learned predictions specifically, see \cite{mitzenmacher2022algorithms}.

Finally, some research has been done on other problems that use predictions quite similar to the item size estimates we deal with in this article:
Azar et al. \cite{azar2021flow} considered a scheduling variant where the size of a job presented upon its revelation is not exact. They designed robust algorithms for the scenario where the amount of distortion is known to the algorithm as well as the scenario where this is not the case \cite{azar2022distortion} and later extended their results to multiple machines \cite{azar2022distortion2}. Another variant of scheduling with estimated job sizes was analyzed by Scully et al. \cite{scully2021uniform}; in their model, however, it is assumed that the actual durations of the jobs are picked randomly and not adversarially. Apart from scheduling, it was recently shown that the simple knapsack problem is exactly $2$-competitive if the algorithm is given almost exact estimates \cite{gehnen2024online,balaban2025online}, and that on general graphs an algorithm cannot achieve a competitive ratio better than the estimate accuracy when giving edge weight estimates in a graph exploration setting \cite{gehnen2025online}.

\subsection{Preliminaries and Notation}

For the sake of this work, we will use the following, standard notation for bin packing problems \cite{coffman2013bin}: 

\begin{definition}[Bin Packing]
	
	In the \textit{Offline Bin Packing} problem an algorithm $A$ receives a list of \textit{items} $L = a_1,...,a_n$ with sizes $c(a_i) \in [0,1]$ for all $i \in \{1,...,n\}$. It then has to find a partition of the items (also called \textit{packing})  into \textit{bins}, such that the sum of the item sizes from each bin does not exceed $1$. The algorithm's goal is to find a packing with as few bins as possible. We define $OPT(L)$ as the minimal number of bins for any possible packing of an input list $L$.
\end{definition}

The online variant of bin packing differs from the offline bin packing, as the number of items and their sizes are not known in the beginning:

\begin{definition}[Online Bin Packing]
	In the \textit{Online Bin Packing} problem the algorithm receives each item $a_i$ together with its size $c(a_i)$. The algorithm needs to decide in which bin the item will be packed before the next item is revealed or the end of the instance is announced.
\end{definition}

In this work, we consider a slightly altered version of the problem described above: 

\begin{definition}
	In the \textit{Online Bin Packing with Estimated Item Sizes} Problem, the algorithm $A$ receives \textit{estimated} sizes $c'(a) \in (0,1]$ for every $a \in L$ as well as the (relative) \textit{accuracy} $\delta \in (0,1]$ in the beginning. Then the items are revealed with their actual sizes $c(a_i)$ and must be designated to some bin before the next item size is revealed. The actual size $c(a)$ of each item is in the interval $[c'(a)(1 - \delta), \min(c'(a)(1 + \delta), 1)]$.
\end{definition}
In other words, we are studying a scenario that one might call "semi-online" since the algorithm knows more about the instance than in the classical online version but the details are still only revealed in an online manner. Note that the case $\delta = 0$ is identical to offline bin packing. The case $\delta = 1$ is not quite the same as regular online bin packing as the algorithm still knows the number of items ahead of the packing; as items can be presented arbitrarily small however, this will not help an algorithm (as an arbitrary large amount of items can be announced, which might just be presented sufficiently small that all fit into one bin).

The packing an algorithm $A$ achieves in an online setting might not be optimal, so we explicitly denote by $A(L)$ the number of bins $A$ needs to pack all items of an input list $L$. The most important metric to determine the quality of an online algorithm is its competitive ratio. As bin packing is a minimization problem, and we are interested in the asymptotic behavior, the following definition is suitable \cite{coffman2013bin}:

\begin{definition}[Competitive Ratio for Minimization Problems]\label{def:regular-cr}
	An algorithm $A$ is called $c$\textit{-competitive} if 
	$$ A(L) \leq c \cdot OPT(L) + K $$
	holds for some constant $K$ and every possible input list $L$.
\end{definition}
For other online problems, an inverse definition for maximization problems or a strict version without the constant $K$ is also used.

\section{Lower Bounds}

Bin packing strategies without information about the input ahead have to pack items rather tightly as they know that the instance could end at any moment. This is not the case for our problem: It might in fact in some cases be wise for an algorithm to use $n$ bins for the first $n$ items if it knows that these bins will reach a sufficient fill level later on. Therefore, the main difficulty such algorithms face is the uncertainty about whether a given subset of items that are yet to be revealed fit together in a single bin. The most challenging input instances are the ones that provide as little information as possible for an algorithm to deal with this uncertainty.

\subsection{Lower Bound of $\frac{4}{3}$ for arbitrary good accuracy}

However, even if the deviation is arbitrarily small, an algorithm cannot achieve a near-optimal result.
This is not particularly surprising, as small deviations already cause situations, in which in the beginning it is hidden if items will fit together or not.
The following construction shows that each algorithm needs to have a competitive ratio of at least $\frac{4}{3}$ by ensuring, that the items, that are placed on already partly filled bins during the first half of the packing process, are the smallest of the entire input instance. 

\begin{theorem}\label{th:general-lower-bound-advanced}
	No bin packing strategy can achieve a competitive ratio of less than $\frac{4}{3}$, even for arbitrary small deviations.
\end{theorem}
\begin{proof}
	We announce a list of items $L$ with $2n$ items with an announced size of $\frac{1}{2}$ for every item in $L$ for some large $n$. The instance consists of two phases; the first one consisting of $\frac{4n}{3}$ items, the second of the remaining $\frac{2n}{3}$ items.
	Items of the first phase will call an item \textit{stacked} if they were placed upon another item by the algorithm, and \textit{laid out} if it they instead were put into a new bin. Each of the first $\frac{4n}{3}$ items will be denoted by $x_{i, k}$ where $i$ and $k$ are essentially counters that track the algorithm's decisions: $i$ equals one plus the number of bins containing two items at the time when $x_{i, k}$ is presented, and $k$ counts the number of items that were revealed after the last item was stacked. (This means that we start counting both $i$ and $k$ from $1$).
	Let $t \leq \frac{2n}{3}$ be the number of items the algorithm stacks, then we observe that for every $i \leq t$ there exists exactly one stacked item which we will call $x_{i, k_i}$. The sizes of the items are
	
$$
		c(x_{1, k}) = \frac{1}{2} - \frac{k\delta}{2n} \text{ and }
		c(x_{i, k}) = c(x_{j, k_{j - 1}}) - \frac{k\delta}{2n^i} 
$$
	
	for $i > 1$, where $x_{j, k_{j - 1}}$ is the last item that was laid out before $x_{i, k}$ was revealed. Note that all items have indeed a size of at least $\frac{1}{2}(1 - \delta)$.
	
	For all $1 \leq i \leq t$ and all $k$ for which $x_{i + 1, k}$ exists we now get
	
	$$ c(x_{i + 1, k}) = c(x_{j, k_{j - 1}}) - \frac{k\delta}{2n^{i + 1}} > c(x_{j, k_{j - 1}}) - \frac{\delta}{2n^i} \geq c(x_{i, k_i}). $$
	
	Furthermore, if $k_i > 1$, we also get
	
	$$ c(x_{i, k_{i - 1}}) > c(x_{i, k_{i - 1}}) - \frac{k\delta}{2n^{i + 1}}  = c(x_{i + 1, k}). $$
	
	Together, these two statements imply that the stacked items $x_{1, k_1},...,x_{t,k_t}$ are smaller than any of the remaining $n - t$ items revealed so far.
	
	The algorithm assigns the first $\frac{4n}{3}$ items into $pn$ bins, where $\frac{2}{3} \leq p \leq \frac{4}{3}$, with $n(\frac{4}{3} - p)$ bins containing two items.
	
In the second phase, $\frac{2n}{3}$ items are presented, again all with size close to $\frac{1}{2}$. Here, $n(\frac{4}{3} - p)$ items of size slightly larger then $\frac{1}{2}$ are presented which will fit only together with the stacked items of the first phase.  The remaining $\frac{2n}{3}-n(\frac{4}{3} - p)$ items will be of size exactly $\frac{1}{2}$.

Therefore, the algorithm will use at least $pn + n(\frac{4}{3} - p) = \frac{4n}{3}$ bins, as all of the the larger items from the second phase will need their own bin in addition to the $pn$ bins of the first phase. The optimal packing however will need only $n$ bins, as the large items of the second phase can be packed with the stacked items of the first phase; and all remaining items are of size at most $\frac{1}{2}$. Therefore all bins of the optimal solution will consist two items.
	
\end{proof}

The idea of ensuring that only the smallest items are stacked was previously also used in lower bound constructions like from Babel et al., that are designed for a setting in which at most two items can be placed in a single bin \cite{babel2004algorithms}.

\subsection{Lower Bound for large prediction errors}\label{sec:large-errors}

Not surprisingly, the more the quality of the estimates degrades, it becomes less likely for an algorithm to achieve a significantly better performance than an algorithm that does not use the given estimates at all. 

When the allowed deviation factor is larger than $\frac{41}{43}$, an online algorithm cannot gain substantial information out of the predictions anymore. 
The following proof uses a modified instance similar to Yao, where a $1.5$ competitive ratio is proven for the classical online bin packing problem without any announcements\cite{yao1980new}.

\begin{theorem}
	No algorithm achieves a competitive ratio better than $1.5$ for a deviation $\delta > \frac{41}{43}$
\end{theorem}
\begin{proof}
	Let $L = L_1L_2L_3$ be the input instance, where each sublist $L_i$ consists of $n$ items for some $n$ divisible by $12$. Each item is announced with size $c'(a) = \frac{43}{168}$ for all $a \in L$. 
	
	The first items of $L_1$ will be presented of size $\frac{1}{7} + \epsilon$, therefore an algorithm can pack one to six of them into the same bin.
	
	As the remaining items of $L_2$ and $L_3$ can be presented as $\frac{43}{168}(1 - \delta) < \frac{1}{84}$, an algorithm is forced to not use more than $\frac{n}{4}$ bins to pack the items of $L_1$. Otherwise, the adversary could decide to present the remaining $L_2$ and $L_3$ sufficiently small such that six items of each of the classes $L_1$, $L_2$, and $L_3$ fit together in one bin. Since the optimal solution then has only used $\frac{n}{6}$ bins, the algorithms solution cannot be better than $1.5$-competitive.
	Therefore, we assume that the algorithm has not taken more than $\frac{n}{4}$ bins for the items of $L_1$.
	
	Next, $n$ items of $L_2$ are presented of size $\frac{1}{3}+\varepsilon$. Again, if the adversary decides to present the items of $L_3$ smaller than $\frac{1}{84}$ each, an optimal solution could consist of $\frac{n}{2}$ used bins (two items of each class, in each of the bins). Therefore, when trying to avoid a competitive ratio of $\frac{3}{2}$ or more, the algorithm can only use up to $\frac{3n}{4}$ bins to pack $L_2$ items on top of the $L_1$ items. 
	Note that only two items of $L_2$ fit together, if zero to two items of $L_1$ are in a bin; and only one item of $L_2$ fits into a bin with three or four items of $L_1$. If more than half of a bin should remain empty, it can fit at most three items of $L_1$ or one item of $L_2$ (possibly combined with a single item of $L_1$).
	As the items could be revealed with the same idea as in \cite{yao1980new}, with analogous calculus it follows that at least $\frac{n}{2}$ bins must be filled by more than $\frac{1}{2}$, given the constraints that the $L_1$ items must completely be contained in $\frac{n}{4}$ bins. 
	
	When finally $n$ $L_3$ items with value $\frac{1}{2}+\varepsilon$ are presented, an algorithm can only place one of them into each bin filled less than $\frac{1}{2}$. As at least $\frac{n}{2}$ bins are already filled by more than half, an algorithm will need at least $1.5n$ bins.
	
	Therefore, no algorithm can achieve a competitive ratio better than $\frac{3}{2}$.
\end{proof}

\section{The Planned Harmonic Algorithm for precise estimates}

Lower bound constructions like Theorem \ref{th:general-lower-bound-advanced} suggest that items with an announced size of around $\frac{1}{2}$ might be the hardest ones for an algorithm to deal with. This is not surprising, as an algorithm cannot know whether two such items fit into the same bin and thus risks leaving a lot of unused space if it makes a wrong decision while packing such an item.

In this section, we present an algorithm called \textsc{Planned-Harmonic} (\textsc{PH}) which achieves a competitive ratio of $\frac{3}{2}$ for rather accurate estimates.

The main idea behind \textsc{PH} is to pack items of size around $\frac{1}{2}$ together with smaller ones and hence avoid such difficult decisions as much as possible. The algorithm $\textsc{Harmonic}_4$ is used as a fallback for items that could not have been packed this way.

\subsection{The Algorithm}

We will call items in the size range between $(\frac{1}{k + 1}, \frac{1}{k}]$ $I_k$-items for $1 \leq k \leq 3$, and $I_4$-items otherwise. Similarly, bins that exclusively contain $I_k$-items are called $I_k$-bins for all $k \leq 4$. The algorithm \textsc{PH} follows the "natural" division into a planning phase which happens before any items are shown, and an update phase which is executed each time an item is revealed.

\subsubsection{Planning Phase}

Before the first item gets revealed, algorithm \ref{alg:ph-plan} is called. Here, \textsc{PH} identifies all items that might be larger than $\frac{1}{2}$ and reserves a separate bin for each of these items. It then assigns items that have a size of at most $\frac{1}{4}$ to those fixed bins in a greedy fashion. 

\begin{algorithm}
	\caption{Planning phase of \textsc{PH}}\label{alg:ph-plan}
	\begin{algorithmic}[1]
		\State $I_{1g} \gets \{a \in L \,\vert\, (1 - \delta)c'(a) > \frac{1}{2} \}$
		\State $I_{1p} \gets \{a \in L \,\vert\, (1 + \delta)c'(a) > \frac{1}{2} \} \setminus I_{1g}$
		\State $I_{4+} \gets \{a \in L \,\vert\, (1 - \delta)c'(a) \leq \frac{1}{4} \}$
		
		\For{$a \in I_{1g}$}
		\State $S \gets$ maximal set in $I_{4+}$ with $(1 + \delta)(c'(a) + \sum_{a' \in S} c'(a')) \leq 1$
		\State $I_{4+} \gets I_{4+} \setminus S$
		\State Plan to pack $a$ and items in $S$ together
		\EndFor
		
		\For{$i \in \{1,...,\vert I_{1p} \vert\}$}
		\State Break if $I_{4+} = \emptyset$
		\State $S_i \gets $ maximal set in $I_{4+}$ with $(1 + \delta)(\frac{1}{2(1 - \delta)} + \sum_{a' \in S_i} c'(a')) \leq 1$
		\State Designate bin $B_i$ for $S_i$
		\EndFor
	\end{algorithmic}
\end{algorithm}

\subsubsection{Update Phase}\label{subsec:update-phase}

Note that items, that are potentially larger than $\frac{1}{2}$ (in the code called $I_{1p}$), are treated differently than items that are guaranteed to be larger than $\frac{1}{2}$ (which are called $I_{1g}$): They were assumed to be as large as possible. The reason is that we strongly prefer $I_1$-items to end up in the reserved bins. 

If an $I_{1p}$-item is revealed to have a size of at most $\frac{1}{2}$ during the update phase, it seems wiser to pack it using the standard harmonic algorithm. This is only sensible as long as there are enough remaining $I_{1p}$-items to be revealed, which then can be used to fill the remaining reserved bins. 

To do so, we assume that every $I_{1p}$-item is as large as possible in the planning phase. This allows us, to decide freely on where to place it after its revelation and is essentially the key reason why \textsc{PH} is $\frac{3}{2}$-competitive. 

The strategy for the update phase is described in Algorithm \ref{alg:ph-update}. Here, $a$ denotes the current item and $B_1,...,B_l$ the bins which already contain an $I_{1p}$-item. With $k$ we denote the number of items from $I_{1p}$ that are assigned into a set $S_i$ in the planning phase.

\begin{algorithm}
	\caption{Update phase of \textsc{PH}}\label{alg:ph-update}
	\begin{algorithmic}[1]
		\If{$a \in I_{1p}$ and $a > \frac{1}{2}$}
		\State Pack $a$ into $B_{l + 1}$, or into a new empty bin if $k = l$
		\ElsIf{$a \in I_{1p}$ and $a \in (\frac{1}{3},\frac{1}{2}]$}
		\State $m \gets $ the number of items in $I_{1p}$ that have not been packed yet
		\If{$m \geq k - l$}
		\State Pack $a$ using \textsc{Harmonic}$_4$
		\Else
		\State Pack $a$ into $B_{l + 1}$
		\EndIf
		\Else
		\State Pack $a$ according to plan, or using \textsc{Harmonic}$_4$ if no plan existed for $a$
		\EndIf
	\end{algorithmic}
\end{algorithm}

In this section, we will see that \textsc{PH} is $\frac{3}{2}$-competitive if the estimates are rather accurate.
\begin{theorem}
	The \textsc{Planned Harmonic} is at most $\frac{3}{2}$-competitive for a deviation $\delta \leq \frac{1}{35}$.
\end{theorem}

To prove this theorem, we start by classifying the $I_1$-items, depending on how the algorithm has dealt with them:

An item of actual size greater than $\frac{1}{2}$ will be called \textit{filled with smaller items} if it is, together with the remaining items of $S_i \subseteq I_{4+}$, packed in the same bin. There are two cases to consider: either all $I_1$-items are filled with smaller items, or not. In the former case, it can easily be seen that all bins except for a constant number have a fill level of at least $\frac{2}{3}$. 

\begin{lemma} 
	\label{lemma:fill-level-b-i}
	For $\delta \leq \frac{1}{35}$ all bins $B_i$ are filled to at least $\frac{2}{3}$, if their respective item $I_{1p}$ is filled with smaller items (except possibly the last one).
\end{lemma}
\begin{proof}
	Each of the bins $B_i$ contains one $I_{1p}$-item as well as all members of the set $S_i$ which was constructed by greedily adding items from $I_{4+}$ while preserving the constraint
	$$ (1 + \delta)\left(\frac{1}{2(1 - \delta)} + c'(S_i)\right) \leq 1 $$
	where $c'(S_i)$ is defined as $\sum_{a' \in S_i} c'(a')$. 
	
	An item in the set $I_{4+}$ has an announced size of at most $\frac{1}{4(1 - \delta)}$. This implies that for the bin $B_i$, the relation
$$
		(1 + \delta)\left(\frac{1}{2(1 - \delta)} + c'(S_i) + \frac{1}{4(1 - \delta)}\right) > 1 \\
		\Leftrightarrow c'(S_i) > \frac{1}{1 + \delta} - \frac{3}{4(1 - \delta)}
$$
	must hold as it would otherwise be possible to augment the set $S_i$ by another $I_{4+}$-item which would contradict the maximality of $S_i$. Since any $I_{1p}$-item has an expected size of more than $\frac{1}{2(1 + \delta)}$, the fill level of $B_i$ has to be at least
$$
		(1 - \delta)(\frac{1}{2(1 + \delta)} + c'(S_i)) \\
		> \frac{3(1 - \delta)}{2(1 + \delta)} - \frac{3}{4} \\
		\geq \frac{2}{3}$$ 
		 for $ \delta \leq \frac{1}{35}$.
\end{proof}
Combining this with the behavior of the underlying harmonic algorithm results in the following observation:
\begin{lemma}
	\label{lemma:all-bins-filled}
	For $\delta \leq \frac{1}{35}$ and if all $I_1$-items can be filled with smaller items, at most $4$ bins do not reach a fill level of at least $\frac{2}{3}$.
\end{lemma}
\begin{proof}
	Lemma \ref{lemma:fill-level-b-i} states that the bins $B_1,...,B_{k-1}$ all have a fill level of at least $\frac{2}{3}$, and the same holds true for all $I_j$-bins except for one for all $j \in \{2,3,4\}$ due to the nature of the algorithm \textsc{Harmonic}$_4$. This leaves us with at most $4$ bins for which we cannot guarantee the desired fill level.
\end{proof}
Using this lemma, we can conclude the proof for the case that all $I_1$-items are paired with smaller items:
\begin{lemma}
	The \textsc{Planned Harmonic} algorithm achieves a competitive ratio of at most $\frac{3}{2}$ if $\delta \leq \frac{1}{35}$ for input instances where all $I_1$-items can be filled with smaller items.
\end{lemma}
\begin{proof}
	If an input list $L$ can be packed into $n$ bins by an optimal packing, all items combined can have at most a size of $n$. As in the situation of Lemma \ref{lemma:all-bins-filled} all but four bins are filled with at least $\frac{2}{3}$ by \textsc{PH}, it follows that all items are be distributed into at most $\frac{3n}{2}+ 4$ bins.
\end{proof}

We now have to deal with the case where some $I_1$-items have not been filled up with smaller items. 
\begin{lemma}
	The \textsc{Planned Harmonic} algorithm achieves a competitive ratio of at most $\frac{3}{2}$ if $\delta \leq \frac{1}{35}$ for input instances when not all $I_{1}$-items got combined with smaller items.
\end{lemma}
\begin{proof} 
	
In this situation, we notice two things: 

First, all items that turned out to be in $[1/3,1/2)$, are considered as $I_2$-items and are packed according to the strategy of a harmonic algorithm (e.g., they are placed together, and not combined with items of other classes). In particular, no such item will be packed into a bin $B_{l+1}$ by line $8$, as the small items are already designated to get packed together with $I_1$ items.

Second, the packing does not contain any $I_4$-bins as all small items ended up being placed in reserved bins.

For the rest of the proof, we can adapt the technique used to prove the competitive ratio of harmonic algorithms. To this end, we define the following weight function $w$ for an item $a$:

$$
w(a)=
\begin{cases}
\frac{1}{k}, c(a) \in (\frac{1}{k + 1}, \frac{1}{k}], 1 \leq k \leq 3 \\
0, c(a) \in (0, \frac{1}{4}] \\
\end{cases}
$$

We now observe that the weight of the whole instance $W(L): =\sum_{a \in L} w(a) \geq \textsc{PH}(L) - 2$ holds, since at most two bins, namely one $I_2$- and one $I_3$-bin each, do not contain items with a combined weight of at least $1$.

Let $\overline{W} = \sup\{\sum_{a \in S} w(a) \,\vert\, S \text{ is a set of items with } \sum_{a \in S} c(a) \leq 1\}$, then we can easily see that $\overline{W} \leq \frac{3}{2}$: This is already the case if $S$ does not contain an $I_1$-item, and if it does contain such an item, at most one more item with a non-zero cost and a weight of at most $\frac{1}{2}$ can fit into the set. As a result, we get 
\begin{align*}
	\textsc{PH}(L) &\leq W(L) + 2 \leq \overline{W} \cdot OPT(L) + 2 \\
	&\leq \frac{3}{2} \cdot OPT(L) + 2.
\end{align*}

This yields the competitive ratio in this remaining case.
\end{proof}
In 2016, an algorithm called \textsc{Reserve-Critical} for online bin packing in the advice model was first introduced by Boyar et al.\cite{boyar2016online}. Since the advice model differs from the setting we are investigating, their \textsc{Reserve-Critical} uses a different strategy. However, their analysis uses a similar approach as \textsc{PH} in this setting with estimates. 

\section{Algorithm for at most two Items per Bin}
One of the main disadvantages of harmonic algorithms is, that items with a size of more than $\frac{1}{2}$ are packed in separate bins, leaving the rest of the bin empty. Therefore, a sensible approach is to combine those items with other items, which are still relatively large.

A simple example, among others, is \textsc{Refined-Harmonic}, which divides the class of large items into several smaller ones, to combine them with smaller items \cite{lee1985simple}. However, even as such approaches promise to slightly improve the competitive ratio of a classic harmonic algorithm, it is of course impossible to break lower bounds like $1.5$.

Additional improvements are possible if adding further constraints: For example, when assuming that a solution might only consist of up to two items per bin, a ratio of $1 + \frac{1}{\sqrt{5}} \approx 1.4472$ is achievable. Thus its competitive ratio significantly undercuts $1.5$, but unfortunately, there remains little room to improve as the currently best known lower bound is $1.4286$ for this particular variant (with previous bounds of $\sqrt{2}$ or $1.4276$)(\cite{babel2004algorithms,fujiwara2015improved, balogh2020online}). Just like \textsc{Refined-Harmonic}, the algorithm of Babel et al. attempts to preserve a ratio between the number of bins with different item configurations \cite{babel2004algorithms}. However, if estimates are available, it can help to overcome the balancing problem between the bins with different configurations. 

One example for instances, when a solution can consist of at most $2$ items are instances, where all items are of size larger than $\frac{1}{3}$. In the setting of online bin packing with items size estimates, one might already see that this condition holds by the given estimates.
\subsection{Delayed-Best-Fit}

To deal with the setting that only two items will fit together in a bin, we present an algorithm that modifies the well-known \textsc{Best-Fit} Algorithm. Remember that \textsc{Best-Fit} packs each item into the fullest bin in that it fits, or packs it in a new bin if it does not fit in any of the used bins:

\begin{algorithm}
	\caption{Delayed Best Fit (DBF)}\label{alg:dbf}
	\begin{algorithmic}[1]
		\State $n \gets $ the number of announced items
		\State $a \gets $ the current item, $c(a) \gets$ the size of the item
		\If{$c(a) \leq \frac{1}{2}$ and $a$ in the first $\frac{1}{3}n$ of items of size smaller $\frac{1}{2}$}
		\If{$a$ fits in a bin $B$ an item of size larger $\frac{1}{2}$}		
		\State Pack $a$ in bin $B$ to the large item
		\Else
		\State Place $a$ in an empty bin
		\EndIf
		\Else
		\State Pack $a$ according to \textsc{Best-Fit}
		\EndIf
	\end{algorithmic}
\end{algorithm}

The first $\frac{1}{3}n$ of items with a size smaller or equal to $\frac{1}{2}$ are also called \emph{special items}, as they are the only ones that are not packed in a Best-Fit manner.

\begin{theorem}
The \textsc{Delayed Best Fit} algorithm achieves a competitive ratio of $\frac{4}{3}$ for all instances where at most two items can be placed into the same bin.
\end{theorem}

To prove this behavior, we first classify the bins used in the analysis:
\begin{figure}
	\centering
	\includegraphics[scale=0.45]{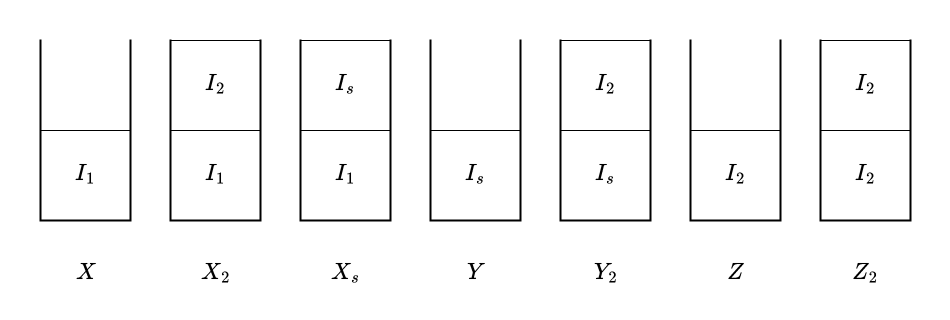}
	\caption{All possible bin configurations in a packing by \textsc{DBF}}\label{fig:bin-configs}
\end{figure}

Figure \ref{fig:bin-configs} depicts all possible bin configurations in a \textsc{DBF}-packing along with their identifiers; later we denote the number of bins of a certain type by their corresponding lowercase letter. Analogously to the analysis of the harmonic algorithm, we call items that are larger than $\frac{1}{2}$ $I_1$, items of size smaller or equal to $\frac{1}{2}$ which are no special item items of type $I_2$, and finally the special items $I_s$.

First,  we need to make the following observations:
\begin{lemma}\label{lemma:no-zs}
	If $y > 0$ in a packing by \textsc{DBF} in an input list $L$, it holds $\textsc{DBF}(L) = n_1 + y + y_2$ with $n_1$ denoting the number of $I_1$ items.
\end{lemma}
\begin{proof}
	\textsc{DBF} packs $I_2$-items using \textsc{Best-Fit} and all special items have a size of at most $\frac{1}{2}$, so before any $I_2$-item is placed into an empty bin it is packed together with a special item. If there exists a $Y$-bin in the packing there can therefore be no bins that only contain $I_2$-items. 
\end{proof}

\begin{lemma}\label{lemma:unfitting-special-items}
	If $y > 0$ in a packing by \textsc{DBF}, then no special item that was packed in a $Y$- or $Y_2$-bin fits together with any $I_1$-item that lies in a $X$- or $X_2$-bin.
\end{lemma}
\begin{proof}
	We first show the statement for special items in $Y$-bins:
	For this, let $a$ be a special item placed in a $Y$-bin and $a'$ an $I_1$-item placed in a $X$- or $X_2$-bin in a packing by \textsc{DBF}. If $a$ was revealed before $a'$ it obviously must have been placed into an empty bin, and as $a'$ was shown, the algorithm must have compared the two items and realized that they do not fit together into the same bin. Note that due to Lemma \ref{lemma:no-zs} there could not have existed any $Z$-bins where the algorithm could have placed $a'$ into. 
	
	If, conversely, $a'$ was revealed before $a$, the two items must have again been compared when $a$ was shown. The bin in which $a'$ was placed cannot have additionally contained a regular $I_2$-item as, per definition, no such item was yet revealed.
	
	For a special item in a $Y_2$-bin we know that it must have been placed there first and the corresponding $I_2$-item was packed into that bin with \textsc{Best-Fit} at a later point. This means that the special item must have been larger than any $I_s$-item that later ended up in a $Y$-bin in the final packing. So when $a'$ is presented, it cannot open a new bin as it is fitted together with all special items in bins of type $Y$, where \textsc{Best-Fit} prefers to place it over opening a new bin. 
\end{proof}

\begin{lemma}\label{lemma:opt-lb}
	With $y' = y +y_2$ and if $y > 0$ and $y + y_2 \geq x_s$ in a packing by \textsc{DBF}, then $OPT(L) \geq n_1 + \frac{y' - x_s}{2}$.
\end{lemma}
\begin{proof}
	From Lemma \ref{lemma:unfitting-special-items} we know that $I_s$-items that were placed in $Y$- or $Y_2$-bins can only fit together in a bin with $I_1$-items that were placed in $X_s$-bins. Only special items contained in $X_s$-bins can fit together with other $I_1$-items, but there must exist $y' - x_s$ special items that cannot have an $I_1$-counterpart in the optimal packing. Thus, every solution must consist of at least $n_1 + \frac{y' - x_s}{2}$ bins.
\end{proof}

With the previous work, we can prove the theorem for the case when $y>0$:

\begin{lemma}
	The \textsc{Delayed Best Fit} algorithm achieves a competitive ratio of $\frac{4}{3}$ for all instances where at most two items can be placed into the same bin and $y>0$.
\end{lemma}

\begin{proof}
	Let $L$ be an arbitrary input instance with $n$ items. 
	We get $\textsc{DBF}(L) = n_1 + y'$ from Lemma \ref{lemma:no-zs} and also have $y'+x_s \leq \frac{n}{3}$. We then can make a case distinction:
	
	\begin{itemize}
		\item $y' \leq x_s$: This implies $y' \leq \frac{1}{6}n$ as $y' + x_s \leq \frac{1}{3}n$. If now $n_1 < \frac{1}{2}n$ we are done as $\textsc{DBF}(L) = n_1 + y' < \frac{1}{2}n + \frac{1}{6}n = \frac{2}{3}n$. Otherwise, we know that $OPT(L) \geq n_1$ and therefore get a competitive ratio of
		
		$$ \frac{n_1 + y'}{n_1} \leq \frac{\frac{1}{2}n + \frac{1}{6}n}{\frac{1}{2}n} = \frac{4}{3}. $$
		
		\item $y' > x_s$: We now use Lemma \ref{lemma:opt-lb} to infer a competitive ratio of
		
		$$
			r \leq \frac{n_1 + y'}{n_1 + \frac{y' - x_s}{2}} = 1 + \frac{x_s + y'}{2n_1 + y' - x_s} 
			\leq 1 + \frac{\frac{1}{3}n}{2n_1 + y' - x_s}
		$$	
		
		for the case. 
		
		If now $n_1 \geq \frac{1}{2}n$ we get $2n_1 + y' - x_s \geq n$ and thus $r \leq \frac{4}{3}$. 
		
		If $n_1 < \frac{1}{2}n$, the inequality $\textsc{DBF}(L) > \frac{2}{3}n$ would imply competitive ratio of $\frac{4}{3}$ as well. If however $\textsc{DBF}(L) \geq \frac{2}{3}n$ holds, then we have

$$
			y' > \frac{1}{6}n + (\frac{1}{2}n - n_1) = \frac{2}{3}n - n_1 \text{ and }
			x_s \leq \frac{1}{3}n - y' < n_1 - \frac{1}{3}n $$  and therefore
$$			y'- x_s > \frac{2}{3}n - n_1 - (n_1 - \frac{1}{3}n) = n - 2n_1.
$$

		But this again means $2n_1 + y' - x_s > n$ and therefore $r < \frac{4}{3}$.
	\end{itemize}
\end{proof}
Finally, we show that the competitive ratio is also achieved in the case $y=0$:
\begin{lemma}
	The \textsc{Delayed Best Fit} algorithm achieves a competitive ratio of $\frac{4}{3}$ for all instances where at most two items can be placed into the same bin and $y=0$.
\end{lemma}
\begin{proof}
	
	Again let $L$ be the instance. It is easy to see that we get an optimal packing if the number of $I_1$ items $n_1 \geq \frac{2}{3}n$, i.e. no regular $I_2$-items were revealed. Otherwise, we know that more than $\frac{1}{3}n$ bins contain two items and there are less than $\frac{1}{3}n$ items left. This means we are done as the packing uses less than $\frac{2}{3}n$ bins.
	
	Note that the desired competitive ratio always follows once we prove that the packing by \textsc{DBF} uses at most $\frac{2}{3}n$ bins as an optimal packing always requires $\frac{1}{2}n$ bins.
\end{proof}
Taking the last two lemmas together, we have seen that \textsc{DBF} achieves a competitive ratio of $\frac{4}{3}$ when at most two items fit into the same bin. We close this section with the remark, that the analysis of \textsc{DBF} is tight and no better competitive ratio can be achieved: 

The lower bound of $\frac{4}{3}$ from Theorem \ref{th:general-lower-bound-advanced} holds for all algorithms and uses at most two items per bin, even for arbitrary accurate estimates.
\section{Final Remarks and Open Questions}

Estimated item sizes add a new dimension to online bin packing: Instead of searching for a single value for the competitive ratio we now want to determine the ratio dependent on the quality of the estimates. Not surprisingly, the problem behaves quite differently depending on the quality of the estimates: While for rather imprecise estimates the problem almost behaves like regular online bin packing, the items can be packed more efficiently, and classical lower bound proofs do not apply anymore if the estimates become more and more precise.

We have established a lower bound of $\frac{4}{3}$ for any $\delta > 0$. Interestingly, the construction used in the proof of this theorem exclusively uses items with an announced size of $\frac{1}{2}$. This suggests that in general, such instances might be the most difficult for an algorithm to handle as they reveal as little information as possible for it to work with in the planning phase.
We have also shown that the proof that establishes a lower bound of $\frac{3}{2}$ for regular online bin packing works as well for sufficiently imprecise estimates. As a result and not surprisingly, we assume that the estimates become essentially useless in those cases. An interesting question however is, at which factor an algorithm can start to ignore those estimates.

We also proposed an algorithm called \textsc{Planned-Harmonic} which achieves a competitive ratio of $1.5$ for all $\delta \leq \frac{1}{35}$. This algorithm follows the simple idea of packing rather large items together with small ones. The analysis shows that either these small items do not take up any additional space or that most bins attain a sufficient fill level. However, \textsc{Planned-Harmonic} struggles with instances where the size of every item lies around $\frac{1}{2}$. This mirrors the observations we made while searching for lower bounds. However, we expect that there are still several ways to improve our algorithm, either by guaranteeing the competitive ratio for less precise estimates or by decreasing its competitive ratio for accurate estimates.

For one restricted case, we presented a strategy called \textsc{Delayed-Best-Fit}, explicitly designed to deal with instances where only two items per bin are possible. It turned out, that adding estimates to restricted bin packing variants also gives an algorithm a significant advantage over strategies without access to those estimates. This setting for classical bin packing is also known as binpacking with cardinality constraints, and it would be interesting to see if similar improvements are also possible when restricting each bin not just to $2$, but also to other numbers of items.
Other variants of the classical bin packing problem include other constraints or features to the model, like occasional repacking, fragmentation, deletion of items, or extending it to more dimensions. It would be interesting to see how and if allowing estimates to algorithms for those variants will be beneficial.

Furthermore, while we have considered a variant where each item value can deviate by some factor from its estimate, other variants for estimates can be seen as realistic. As an example, if the deviation can be an additive constant, this can model situations where the actual size cannot be saved due to e.g. rounding errors or space constraints when saving them. While it is likely that this will not change the behavior if all items are relatively large, for small items it allows the adversary to completely remove them. However, we expect that results mostly translate between an additive and multiplicative setting, as the crucial information, which might get lost, is, if an item is larger or smaller than a threshold like $\frac{1}{2}$.
For now, we require that the given estimates are true, so all actual item sizes actually ly within the tolerance of the estimate. A reasonable relaxation would be to consider a setting, where e.g. a limited number of values is allowed to deviate further from the estimate.

Finally, we expect that insights into this variant will help in understanding other common settings and vice versa, for example as the information an algorithm deducts from the estimates could also be given to it in an advice setting.
\bibliography{bibliography.bib}

\begin{thebibliography}{10}

\bibitem{angelopoulos2018online}
Spyros Angelopoulos, Christoph D{\"u}rr, Shahin Kamali, Marc~P Renault, and Adi
  Ros{\'e}n.
\newblock Online bin packing with advice of small size.
\newblock {\em Theory of Computing Systems}, 62:2006--2034, 2018.

\bibitem{angelopoulos2021online}
Spyros Angelopoulos, Shahin Kamali, and Kimia Shadkami.
\newblock Online bin packing with predictions.
\newblock {\em International Joint Conference of Artificial Intelligence
  (AJCAI)}, 2022.

\bibitem{azar2021flow}
Yossi Azar, Stefano Leonardi, and Noam Touitou.
\newblock Flow time scheduling with uncertain processing time.
\newblock In {\em Proceedings of the 53rd Annual ACM SIGACT Symposium on Theory
  of Computing}, pages 1070--1080, 2021.

\bibitem{azar2022distortion}
Yossi Azar, Stefano Leonardi, and Noam Touitou.
\newblock Distortion-oblivious algorithms for minimizing flow time.
\newblock In {\em Proceedings of the 2022 Annual ACM-SIAM Symposium on Discrete
  Algorithms (SODA)}, pages 252--274. SIAM, 2022.

\bibitem{azar2022distortion2}
Yossi Azar, Eldad Peretz, and Noam Touitou.
\newblock Distortion-oblivious algorithms for scheduling on multiple machines.
\newblock In {\em 33rd International Symposium on Algorithms and Computation
  (ISAAC 2022)}. Schloss Dagstuhl-Leibniz-Zentrum f{\"u}r Informatik, 2022.

\bibitem{babel2004algorithms}
Luitpold Babel, Bo~Chen, Hans Kellerer, and Vladimir Kotov.
\newblock Algorithms for on-line bin-packing problems with cardinality
  constraints.
\newblock {\em Discrete Applied Mathematics}, 143(1-3):238--251, 2004.

\bibitem{balaban2025online}
Jakub Balabán, Matthias Gehnen, Henri Lotze, Finn Seesemann, and Moritz
  Stocker.
\newblock Online knapsack problems with estimates.
\newblock {\em arXiv preprint arXiv:2504.21750}, 2025.

\bibitem{balogh2017new}
J{\'a}nos Balogh, J{\'o}zsef B{\'e}k{\'e}si, Gy{\"o}rgy D{\'o}sa, Leah Epstein,
  and Asaf Levin.
\newblock A new and improved algorithm for online bin packing.
\newblock {\em 26th Annual European Symposium on Algorithms (ESA 2018)}, 2018.

\bibitem{balogh2020online}
J{\'a}nos Balogh, J{\'o}zsef B{\'e}k{\'e}si, Gy{\"o}rgy D{\'o}sa, Leah Epstein,
  and Asaf Levin.
\newblock Online bin packing with cardinality constraints resolved.
\newblock {\em Journal of Computer and System Sciences}, 112:34--49, 2020.

\bibitem{balogh2021new}
J{\'a}nos Balogh, J{\'o}zsef B{\'e}k{\'e}si, Gy{\"o}rgy D{\'o}sa, Leah Epstein,
  and Asaf Levin.
\newblock A new lower bound for classic online bin packing.
\newblock {\em Algorithmica}, 83:2047--2062, 2021.

\bibitem{balogh2012new}
J{\'a}nos Balogh, J{\'o}zsef B{\'e}k{\'e}si, and G{\'a}bor Galambos.
\newblock New lower bounds for certain classes of bin packing algorithms.
\newblock {\em Theoretical Computer Science}, 440:1--13, 2012.

\bibitem{borodin2005online}
Allan Borodin and Ran El-Yaniv.
\newblock {\em Online computation and competitive analysis}.
\newblock Cambridge University Press, 2005.

\bibitem{boyar2017online}
Joan Boyar, Lene~M Favrholdt, Christian Kudahl, Kim~S Larsen, and Jesper~W
  Mikkelsen.
\newblock Online algorithms with advice: a survey.
\newblock {\em ACM Computing Surveys (CSUR)}, 50(2):1--34, 2017.

\bibitem{boyar2016online}
Joan Boyar, Shahin Kamali, Kim~S Larsen, and Alejandro L{\'o}pez-Ortiz.
\newblock Online bin packing with advice.
\newblock {\em Algorithmica}, 74:507--527, 2016.

\bibitem{brown1979lower}
Donna~J Brown.
\newblock A lower bound for on-line one-dimensional bin packing algorithms.
\newblock Technical report, Technical Report R-864, Coordinated Sci. Lab.,
  Urbana, Illinois, 1979.

\bibitem{coffman2013bin}
Edward~G Coffman, J{\'a}nos Csirik, G{\'a}bor Galambos, Silvano Martello,
  Daniele Vigo, et~al.
\newblock Bin packing approximation algorithms: Survey and classification.
\newblock In {\em Handbook of combinatorial optimization}, pages 455--531.
  Springer, 2013.

\bibitem{fernandez1981bin}
W~Fernandez~de La~Vega and George~S. Lueker.
\newblock Bin packing can be solved within 1+ $\varepsilon$ in linear time.
\newblock {\em Combinatorica}, 1(4):349--355, 1981.

\bibitem{fujiwara2015improved}
Hiroshi Fujiwara and Koji Kobayashi.
\newblock Improved lower bounds for the online bin packing problem with
  cardinality constraints.
\newblock {\em Journal of Combinatorial Optimization}, 29(1):67--87, 2015.

\bibitem{gehnen2025online}
Matthias Gehnen, Ralf Klasing, and Émile Naquin.
\newblock Graph exploration with edge weight estimates.
\newblock {\em arXiv preprint arXiv:2501.18496}, 2025.

\bibitem{gehnen2024online}
Matthias Gehnen, Henri Lotze, and Peter Rossmanith.
\newblock {Online Simple Knapsack with Bounded Predictions}.
\newblock In {\em 41st International Symposium on Theoretical Aspects of
  Computer Science (STACS 2024)}, volume 289, 2024.

\bibitem{hoberg2017logarithmic}
Rebecca Hoberg and Thomas Rothvoss.
\newblock A logarithmic additive integrality gap for bin packing.
\newblock In {\em Proceedings of the Twenty-Eighth Annual ACM-SIAM Symposium on
  Discrete Algorithms}, pages 2616--2625. SIAM, 2017.

\bibitem{johnson1973near}
David~S Johnson.
\newblock {\em Near-optimal bin packing algorithms}.
\newblock PhD thesis, Massachusetts Institute of Technology, 1973.

\bibitem{johnson1974worst}
David~S. Johnson, Alan Demers, Jeffrey~D. Ullman, Michael~R Garey, and
  Ronald~L. Graham.
\newblock Worst-case performance bounds for simple one-dimensional packing
  algorithms.
\newblock {\em SIAM Journal on computing}, 3(4):299--325, 1974.

\bibitem{kamali2020online}
Shahin Kamali.
\newblock Online bin packing with predictions.
\newblock 2020.

\bibitem{komm2016introduction}
Dennis Komm.
\newblock {\em Introduction to Online Computation}.
\newblock Springer, 2016.

\bibitem{lee1985simple}
Chan~C Lee and Der-Tsai Lee.
\newblock A simple on-line bin-packing algorithm.
\newblock {\em Journal of the ACM (JACM)}, 32(3):562--572, 1985.

\bibitem{liang1980lower}
Frank~M Liang.
\newblock A lower bound for on-line bin packing.
\newblock {\em Information processing letters}, 10(2):76--79, 1980.

\bibitem{mikkelsen2015randomization}
Jesper~W Mikkelsen.
\newblock Randomization can be as helpful as a glimpse of the future in online
  computation.
\newblock {\em 43rd International Colloquium on Automata, Languages, and
  Programming (ICALP 2016)}, 2016.

\bibitem{mitzenmacher2022algorithms}
Michael Mitzenmacher and Sergei Vassilvitskii.
\newblock Algorithms with predictions.
\newblock {\em Communications of the ACM}, 65(7):33--35, 2022.

\bibitem{purohit08}
Manish Purohit, Zoya Svitkina, and Ravi Kumar.
\newblock Improving online algorithms via ml predictions.
\newblock In S.~Bengio, H.~Wallach, H.~Larochelle, K.~Grauman, N.~Cesa-Bianchi,
  and R.~Garnett, editors, {\em Advances in Neural Information Processing
  Systems}, volume~31. Curran Associates, Inc., 2018.

\bibitem{ramanan1989line}
Prakash Ramanan, Donna~J Brown, Chung-Chieh Lee, and Der-Tsai Lee.
\newblock On-line bin packing in linear time.
\newblock {\em Journal of Algorithms}, 10(3):305--326, 1989.

\bibitem{renault2015online}
Marc~P Renault, Adi Ros{\'e}n, and Rob van Stee.
\newblock Online algorithms with advice for bin packing and scheduling
  problems.
\newblock {\em Theoretical Computer Science}, 600:155--170, 2015.

\bibitem{scully2021uniform}
Ziv Scully, Isaac Grosof, and Michael Mitzenmacher.
\newblock Uniform bounds for scheduling with job size estimates.
\newblock {\em 13th Innovations in Theoretical Computer Science Conference
  (ITCS 2022)}, 2022.

\bibitem{seiden2002online}
Steven~S Seiden.
\newblock On the online bin packing problem.
\newblock {\em Journal of the ACM (JACM)}, 49(5):640--671, 2002.

\bibitem{10.1145/2786.2793}
Daniel~D. Sleator and Robert~E. Tarjan.
\newblock Amortized efficiency of list update and paging rules.
\newblock {\em Commun. ACM}, 28(2):202–208, feb 1985.
\newblock URL: \url{https://doi.org/10.1145/2786.2793}.

\bibitem{van1992improved}
Andr{\'e} van Vliet.
\newblock An improved lower bound for on-line bin packing algorithms.
\newblock {\em Information processing letters}, 43(5):277--284, 1992.

\bibitem{yao1980new}
Andrew Chi-Chih Yao.
\newblock New algorithms for bin packing.
\newblock {\em Journal of the ACM (JACM)}, 27(2):207--227, 1980.

\end{thebibliography}
\appendix
\end{document}